\documentclass[12pt]{article}
\usepackage{graphicx}
\usepackage{amsmath}
\usepackage{amsbsy}
\usepackage{amsfonts}
\usepackage{amsthm}
\begin{document}

\title{Q-stars in extra dimensions}

\author{Athanasios Prikas}

\date{}

\maketitle

Physics Department, National Technical University, Zografou
Campus, 157 80 Athens, Greece.\footnote{e-mail:
aprikas@central.ntua.gr}

\begin{abstract}
We study q-stars with global and local $U(1)$ symmetry in extra
dimensions in asymptotically anti de Sitter or flat spacetime.
The behavior of the mass, radius and particle number of the star
is quite different in 3 dimensions, but in 5, 6, 8 and 11
dimensions is similar to the behavior in 4.
\end{abstract}

PACS number(s): 11.27.+d, 04.40.-b

\newpage

\section{Introduction}

Boson stars are stable field configurations of a massive complex
scalar field coupled to gravity,
\cite{boson-stars-old1,boson-stars-old2,boson-stars-new}. Quartic
self-interactions, \cite{boson-stars-quartic}, and the case of the
charged scalar field, \cite{boson-stars-charged}, have also been
investigated. Gravity stabilizes the star, preventing it form
decay into free particles.

Another class of boson stars are the non-topological soliton
stars. They are stable, even in the absence of gravity. They
appear as relativistic generalizations,
\cite{large-soliton-stars1,large-soliton-stars2,large-soliton-stars3}
of non-topological solitons, \cite{nontop-solitons}, or q-balls.
Q-balls is a subclass of non-topological solitons in Lagrangians
with a global $U(1)$, \cite{qballs-initial}, or $SU(3)$, $SO(3)$
symmetry, \cite{qballs-nonabelian}, or with a local $U(1)$
symmetry, \cite{qballs-charged}. There are q-stars with one or
two scalar fields, \cite{qstars-global}, q-stars with non-abelian
symmetries, \cite{qstars-nonabelian}, with fermions and scalars,
\cite{qstars-fermion}, and with a local $U(1)$ symmetry,
\cite{qstars-charged} in asymptotically flat or anti de Sitter
spacetime, \cite{qstars-AdS}. Recently, a great amount of
theoretical interest has been focused on anti de Sitter
spacetime, in any dimension, due to the close relation between
gravitating fields within an anti de Sitter spacetime and a field
theory at the boundary of the above spacetime,
\cite{witten,maldacena,string review}. Boson (not soliton) stars
with negative cosmological constant and in 3, 4, 5 and 6
dimensions are investigated in \cite{boson-stars-extradim}.

The purpose of the present work is to study the formation of
q-stars, with global or local $U(1)$ symmetry in 5, 6, 8 and 11
dimensions. We also include results in 4 dimensions for
comparison, and in 3 for completeness. We investigate their
properties, mass, radius, particle number and the value of the
scalar field at the center of the star and study their
differences resulting form the different dimensionality or the
effect of a negative cosmological constant. We also study the
stability of the q-star with respect to gravitational collapse
and to fission into free particles.

\section{Q-stars in any dimension}

We consider a static spherically symmetric metric:
\begin{equation}\label{1}
ds^2=-e^{\nu}dt^2+e^{\lambda}d\rho^2+\rho^2d\Omega_{D-2}^2\ ,
\end{equation}
with $g_{tt}=-e^{\nu}$, $D$ the spacetime dimensionality and
$d\Omega_{D-2}^2$ the line element for a $(D-2)$-dimensional unit
sphere. The action for a scalar field coupled to gravity in $D$
dimensions is:
\begin{equation}\label{2}
S_D=-\int_\mathcal{S}d^Dx\left[\frac{\sqrt{-g_D}(R-2\Lambda)}{16\pi
G_D}+L_{\textrm{matter}}\right]-\frac{1}{8\pi
G_D}\int_{\partial\mathcal{S}}d^{D-1}x\sqrt{-h}K\ ,
\end{equation}
where the second term is the Hawking-Gibbons term,
\cite{gibbons}, $\mathcal{S}$ is the spacetime region with
$\partial\mathcal{S}$ its boundary, $R$ and $K$ are the traces of
the curvature and $\Lambda$ stands for the cosmological constant.
The matter Lagrangian is:
\begin{equation}\label{3}
L_{\textrm{matter}}=(\partial_{\mu}\phi)^{\ast}(\partial^{\nu}\phi)-U\
.
\end{equation}
The Einstein and Lagrange equations are respectively:
\begin{equation}\label{4}
G_{\mu}^{\ \nu}\equiv R_{\mu}^{\ \nu}-\frac{1}{2}\delta_{\mu}^{\
\nu}=8\pi G_DT_{\mu}^{\ \nu}-\Lambda\delta_{\mu}^{\ \nu}\ ,
\end{equation}
\begin{equation}\label{5}
\phi_{;\lambda}^{\hspace{1em};\lambda}-\frac{dU}{d|\phi|^2}\phi=0
\ ,
\end{equation}
with $G_{\mu\nu}$ the Einstein tensor and $T_{\mu\nu}$ the
energy-momentum tensor:
\begin{equation}\label{6}
T_{\mu\nu}={({\partial}_{\mu}\phi)}^{\ast}({\partial}_{\nu}\phi)+
({\partial}_{\mu}\phi){({\partial}_{\nu}\phi)}^{\ast}
-g_{\mu\nu}[g^{\alpha\beta}{({\partial}_{\alpha}\phi)}^{\ast}({\partial}_{\beta}\phi)]
-g_{\mu\nu}U\ .
\end{equation}

We will now insert the q-soliton ansatz writing:
\begin{equation}\label{7}
\phi(\vec{\rho},t)=\sigma(\rho)e^{-\imath\omega t} \ .
\end{equation}
with $\omega$ the eigen-frequency with which the star rotates
within its internal $U(1)$ space. We define:
\begin{equation}\label{8} A=e^{-\lambda}\ ,
\hspace{1em} B=e^{-\nu}\ ,
\end{equation}
\begin{equation}\label{9}
\begin{split}
W\equiv e^{-\nu}{\left(\frac{\partial\phi}{\partial
t}\right)}^{\ast}\left(\frac{\partial\phi}{\partial t}\right)=
e^{-\nu}{\omega}^2{\sigma}^2\ , \\ V\equiv
e^{-\lambda}{\left(\frac{\partial\phi}{\partial\rho}\right)}^{\ast}
\left(\frac{\partial\phi}{\partial\rho}\right)=
e^{-\lambda}{\sigma'}^2
\end{split}
\end{equation}
and rescale:
\begin{equation}\label{10}
\begin{split}
\tilde{\rho}=\rho m\ , \hspace{1em} \tilde{\omega}&=\omega/m\ ,
\hspace{1em} \tilde{\phi}=\phi/m^{\frac{D-2}{2}} \ , \\
\widetilde{U}=U/m^D\ , \hspace{1em} \widetilde{W}&=W/m^D\ ,
\hspace{1em} \widetilde{V}=V/m^D\ .
\end{split}
\end{equation}
Gravity is important when $R\sim G_DM(R)$, where $M(R)$ is the
mass within a sphere of radius $R$. For a q-star $U\sim W\sim
m^D$, $V\sim\epsilon^2m^D$ with $\epsilon$:
$$\epsilon\equiv\sqrt{8\pi Gm^{D-2}}\ .$$ This is a very small
quantity for $m$ of the order of magnitude of some (hundreds)
$GeV$. Quantities of the same order of magnitude as $\epsilon$
will be neglected. We also redefine:
\begin{equation}\label{11}
\tilde{r}=\epsilon\tilde{\rho}, \hspace{1em}
\widetilde{\Lambda}=\frac{\Lambda}{8\pi Gm^D}\ .
\end{equation}
As we will see, $\widetilde{\Lambda}$ may vary from zero, down to
a $\sim-1$ value. This means that the cosmological constant
$\Lambda$ has the same order of magnitude as:
$$\Lambda\sim-Gm^D\ .$$ We use a (rescaled) potential admitting q-ball type solutions in
the absence of gravity, namely:
\begin{equation}\label{12}
\widetilde{U}=|\tilde{\phi}|^2\left(1-|\tilde{\phi}|^2+\frac{1}{3}|\tilde{\phi}|^4\right)
=\tilde{\sigma}^2\left(1-\tilde{\sigma}^2+\frac{1}{3}\tilde{\sigma}^4\right)\
.
\end{equation}
Dropping form now on the tildes and the $O(\epsilon)$ quantities
form the Lagrange equation we find an analytical solution for the
matter scalar:
\begin{equation}\label{13}
\sigma=(1+\omega B^{1/2})^{1/2}\ .
\end{equation}

We will now find the eigenvalue equation for the frequency.
Within the interior of the star both the matter end metric fields
vary very slowly, because the radial derivative is of order of
$O(\sqrt{\epsilon})$. Within the star surface, the metric fields
vary slowly but the matter scalar rapidly. Dropping the
$O(\epsilon)$ terms from the Lagrange equation within the surface
we take:
\begin{equation}\label{14}
V+W-U=0\ .
\end{equation}
At the inner edge of the surface $\sigma'$ is zero in order to
match the interior with the surface solution. So, at the inner
edge of the surface the equality $W=U$ together with eq. \ref{13}
gives:
\begin{equation}\label{15}
\omega=\frac{A_{\textrm{sur}}^{1/2}}{2}=\frac{B_{\textrm{sur}}^{-1/2}}{2}\
,
\end{equation}
where $A_{\textrm{sur}}$, $B_{\textrm{sur}}$ denote the value of
the metrics at the surface of the star. Eq. \ref{15} is the
eigenvalue equation for the frequency of the q-star and has the
right limiting value ($\omega=1/2$) in the absence of gravity
($A_{\textrm{sur}}=1$), when the potential is given by eq.
\ref{12}.

We will now turn to the Einstein equations. With the rescalings
and redefinitions of eqs. \ref{8}-\ref{11} and dropping the
$O(\epsilon)$ terms, they take the simple form:
\begin{equation}\label{16}
\frac{A-1}{2r^2}(D-3)(D-2)+\frac{A'}{2r}(D-2)=-W-U-\Lambda\ ,
\end{equation}
\begin{equation}\label{17}
\frac{A-1}{2r^2}(D-3)(D-2)-(D-2)\frac{A}{2r}\frac{B'}{B}=W-U-\Lambda\
.
\end{equation}

The total mass of the field configuration can be estimated by the
$-T_0^{\ 0}$ component of the energy-momentum tensor:
\begin{equation}\label{18}
M=\frac{2\pi^{(D-1)/2}}{\Gamma\left(\frac{D-1}{2}\right)}\int_0^Rdr(U+W)r^{D-2}\
,
\end{equation}
with $R$ the star radius, defined by the Schwarzschild condition
$A(r)=1/B(r)$. There is another relation connecting the metrics
with the mass trapped within a sphere of radius $\rho$, namely:
\begin{equation}\label{19}
A(\rho)=1-\frac{2Gm_{\rho}}{r^{D-3}}-\frac{2\Lambda\rho^2}{(D-2)(D-1)}\
.
\end{equation}
$m_{\rho}$ is straightforward connected to the total mass $M$:
\begin{equation}\label{20}
M=\frac{D-2}{8\pi}\frac{2\pi^{(D-1)/2}}{\Gamma\left(\frac{D-1}{2}\right)}m_{\rho}\
,\hspace{1em}\rho\rightarrow\infty\ ,
\end{equation}
which, with our rescalings, takes the form:
\begin{equation}\label{21}
M=(D-2)\frac{\pi^{(D-1)/2}}{\Gamma\left(\frac{D-1}{2}\right)}r^{D-3}\left(1-A(r)-
\frac{2\Lambda r^2}{(D-2)(D-1)}\right)\
,\hspace{1em}r\rightarrow\infty\ .
\end{equation}
There is a Noether current:
\begin{equation}\label{22}
j^{\mu}=\imath
g^{\mu\nu}(\phi\partial_{\nu}\phi^{\ast}-\phi^{\ast}\partial_{\nu}\phi)
\end{equation}
which gives a conserved Noether charge:
\begin{equation}\label{23}
N=\int_0^{\infty}d^{D-1}x\sqrt{-g_D}j^t=
\frac{4\pi^{(D-1)/2}}{\Gamma\left(\frac{D-1}{2}\right)}\int_0^Rdr\omega\sigma^2r^{D-2}
\sqrt{\frac{B}{A}}\ .
\end{equation}
The star owes its stability to the conserved Noether charge. The
field configuration is stable with respect to fission into free
particles when the star mass is less than the total energy of the
free particles with the same charge.

The Einstein equations have no analytical solution within the
star, except for the $D=3$, $\Lambda=0$ case:
\begin{equation}\label{24}
A(r)=1-\frac{3}{4}r^2\ ,\hspace{1em}B(r)=\frac{1}{4\omega^2}\ ,
\end{equation}
\begin{equation}\label{25}
M=\pi(1-4\omega^2)\ ,\hspace{1em}N=4\pi(1-2\omega)\ .
\end{equation}
Every quantity (total mass, e.t.c.) is $D$-dependent according to
the eqs. \ref{10}-\ref{11}, but we use one figure for different
dimensions for brevity.

\begin{figure}
\centering
\includegraphics{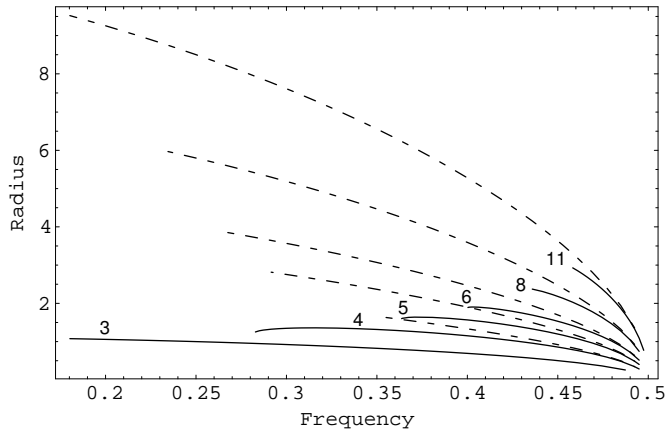}
\caption{The radius of a q-star as a function of its frequency in
an asymptotically flat spacetime. The numbers within the figures
denote the spacetime dimensionality. Dashed lines correspond to
charged q-stars with $e=1$. Figures \ref{figure1}-\ref{figure7}
refer to asymptotically flat spacetime. We do not consider
charged q-stars in $2+1$ dimensions. When lowering $\omega$ for a
global $U(1)$ symmetry or $\theta_{\textrm{sur}}$ for the local
case, or, equivalently, $A_{\textrm{sur}}$ for any case, the star
radius is in generally larger, because stronger surface gravity is
generated by larger solitons. Charged stars have larger radii due
to the electrostatic repulsion between the different parts of the
star.} \label{figure1}
\end{figure}

\begin{figure}
\centering
\includegraphics{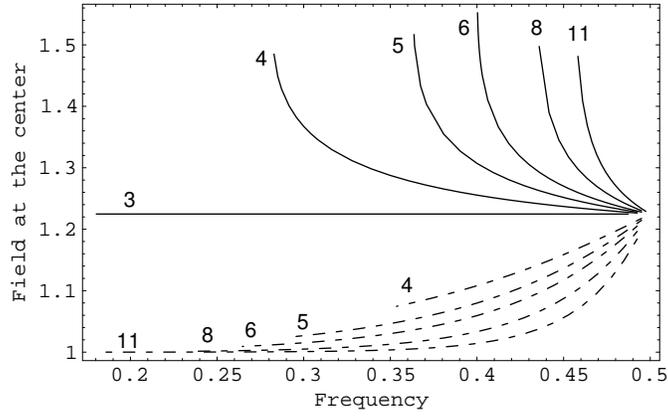}
\caption{The value of the scalar field at the center of the q-star
as a function of its frequency. Dashed lines correspond to charged
q-stars with $e=1$. In $2+1$ dimensions eqs. \ref{13}, \ref{15}
and \ref{24} give $\sigma(r)^2=1.5$} \label{figure2}
\end{figure}

\begin{figure}
\centering
\includegraphics{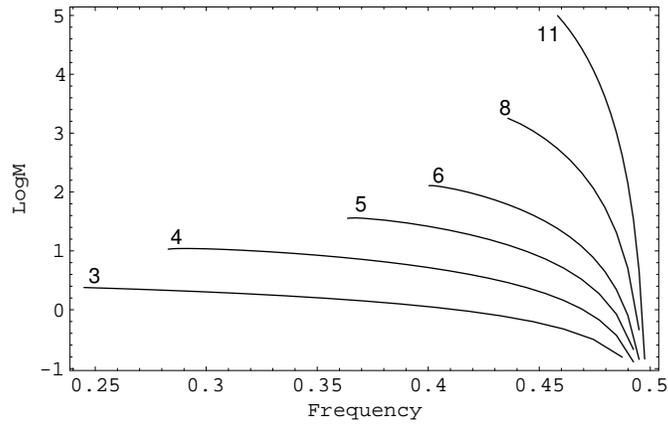}
\caption{The mass of a q-star with global symmetry as a function
of its frequency.} \label{figure3}
\end{figure}

\begin{figure}
\centering
\includegraphics{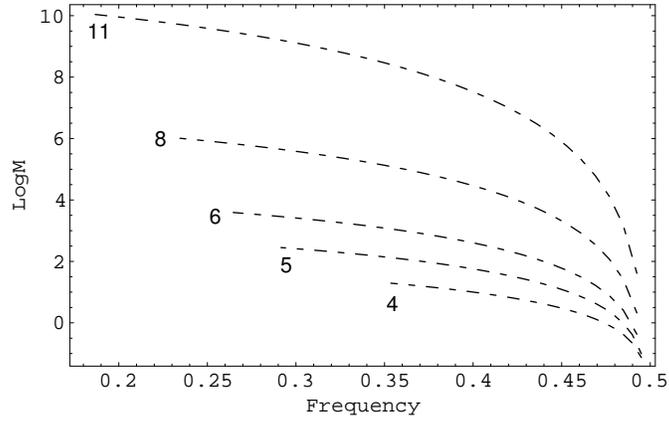}
\caption{The mass of a charged q-star as a function of
$\theta_{\textrm{sur}}$ with $e=1$.} \label{figure4}
\end{figure}

\begin{figure}
\centering
\includegraphics{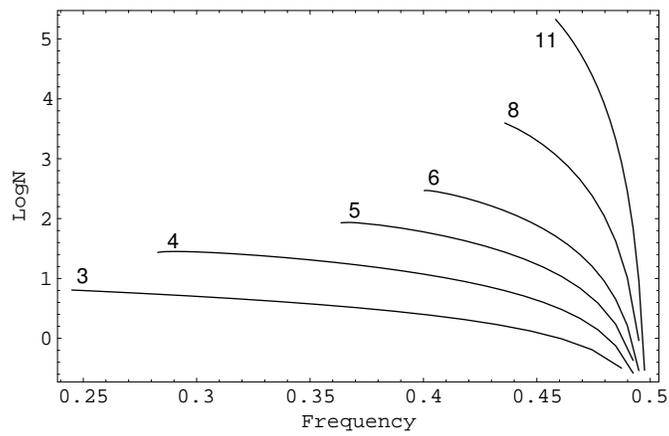}
\caption{The particle number of a q-star with global symmetry as
a function of its frequency.} \label{figure5}
\end{figure}

\begin{figure}
\centering
\includegraphics{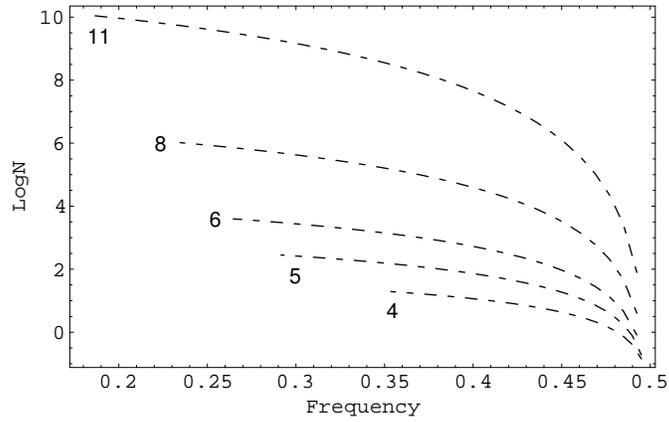}
\caption{The particle number of a charged q-star with $e=1$ as a
function of $\theta_{\textrm{sur}}$.} \label{figure6}
\end{figure}

\begin{figure}
\centering
\includegraphics{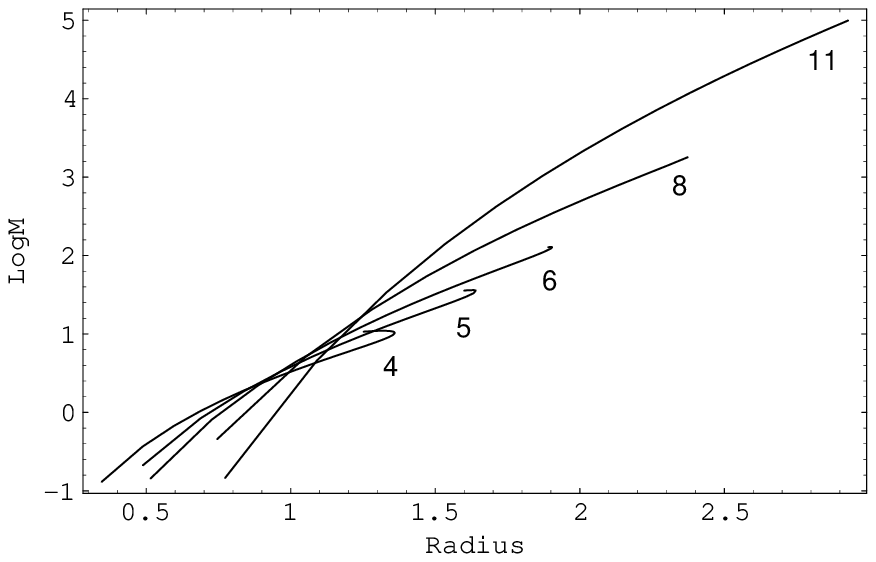}
\caption{The mass of a q-star with global symmetry as a function
of its radius.} \label{figure7}
\end{figure}

\section{The case of local symmetry}

We will now investigate the case of a q-star with a local $U(1)$
symmetry, i.e. a charged q-star. Charged q-balls,
\cite{qballs-charged}, charged q-stars, \cite{qstars-charged} and
boson (not solitonic) stars with local $U(1)$ symmetry,
\cite{boson-stars-charged}, \cite{bosonstar-scalartensor-charged},
have been considered. The difference between the global case is
the form that the matter Lagrangian takes:
\begin{equation}\label{c1}
L_{\textrm{matter}}=g^{\mu\nu}(D_{\mu}\phi)^{\ast}(D_{\nu}\phi)-U-\frac{1}{4}
F^{\mu\nu}F_{\mu\nu}\ ,
\end{equation}
with:
$$F_{\mu\nu}\equiv\partial_{\mu}A_{\nu}-\partial_{\nu}A_{\mu}\
,\hspace{1em} D_{\mu}\phi\equiv\partial_{\mu}\phi-\imath
eA_{\mu}\ .$$ $e$ is the charge or field strength and $A_{\mu}$
is the gauge field. The energy-momentum tensor is:
\begin{eqnarray}\label{c2}
T_{\mu\nu}=(D_{\mu}\phi)^{\ast}(D_{\mu}\phi)+
(D_{\mu}\phi)(D_{\mu}\phi)^{\ast}-
g_{\mu\nu}[g^{\alpha\beta}(D_{\alpha}\phi)^{\ast}(D_{\beta}\phi)]
\nonumber\\
-g_{\mu\nu}U-\frac{1}{4}g_{\mu\nu}F^{\alpha\beta}F_{\alpha\beta}+g^{\alpha\beta}
F_{\nu\alpha}F_{\nu\beta}\ ,
\end{eqnarray}
and the equation of motion for the matter scalar field is:
\begin{equation}\label{c3}
\left[\frac{1}{\sqrt{-g}}D_{\mu}(\sqrt{-g}g^{\mu\nu}D_{\mu})-\frac{dU}{d|\phi|^2}\right]\phi=0\
.
\end{equation}

In order to realize a static metric we choose
$A_{\mu}=(A_0,0,0,0)$, i.e.: we eliminate any magnetic fields. It
is very useful to define a new quantity:
\begin{equation}\label{c4}
\theta=\omega+eA_0\ ,
\end{equation}
which is rescaled in the same way as the frequency. We use the
same ansatz for the scalar field as in the case of global
symmetry and rescale as in eqs. \ref{10}, \ref{11}. The rescaling
for the charge is similar to the one used in charged boson stars:
\begin{equation}\label{c5}
\tilde{e}=e{\epsilon}^{-1}\ .
\end{equation}
Dropping the $O(\epsilon)$ quantities and the tildes from the
Lagrange equation for the scalar field, we find for the interior:
\begin{equation}\label{c6}
\sigma=(1+\theta B^{1/2})^{1/2}\ .
\end{equation}
Repeating the discussion of eqs. \ref{14}-\ref{15} we find:
\begin{equation}\label{c7}
\theta_{\textrm{sur}}=\frac{A_{\textrm{sur}}^{1/2}}{2}=\frac{B_{\textrm{sur}}^{-1/2}}{2}\
.
\end{equation}
The new dynamical quantity $\theta$ plays now the role of the
frequency and eq. \ref{c7} is the eigenvalue equation for this
``frequency".

The Einstein equations take now the form:
\begin{equation}\label{c8}
\frac{A-1}{2r^2}(D-3)(D-2)+\frac{A'}{2r}(D-2)=-W-U-\frac{\theta'^2}{2e^2}-\Lambda\
,
\end{equation}
\begin{equation}\label{c9}
\frac{A-1}{2r^2}(D-3)(D-2)-(D-2)\frac{A}{2r}\frac{B'}{B}=W-U-\frac{\theta'^2}{2e^2}-\Lambda\
,
\end{equation}
and the equation of motion for the gauge field:
\begin{equation}\label{c10}
\theta''+\left[\frac{D-2}{r}+\frac{1}{2}\left(\frac{A'}{A}+\frac{B'}{B}\right)\right]\theta'
-\frac{2e^2\sigma^2\theta}{A}=0\ .
\end{equation}
At the center of the star the electric field is zero, so
$\theta'(0)=0$. The absence of electric fields at infinity leads
to $\theta(\infty)=\omega$. We numerically solve the coupled
system of \ref{c8}-\ref{c10}. For the exterior of the star we
find the analytical solution:
\begin{equation}\label{c11}
\theta'=\frac{\Gamma\left(\frac{D-1}{2}\right)}{2\pi^{\frac{D-1}{2}}}e^2Q\
,
\end{equation}
and the contribution of the exterior to the total energy is:
\begin{equation}\label{c12}
M_{\textrm{ext.}}=\frac{e^2Q^2}{2}
\frac{\Gamma\left(\frac{D-1}{2}\right)}{2\pi^{\frac{D-1}{2}}}
\int_R^{\infty}dr\frac{1}{r^{D-2}}\ ,
\end{equation}
with $R$ the radius of the star.

For a charged q-star, the Noether current is:
\begin{equation}\label{c13}
j^{\mu}=\imath g^{\mu\nu}(\phi
D_{\nu}\phi^{\ast}-\phi^{\ast}D_{\nu}\phi)\ ,
\end{equation}
and the corresponding particle number:
\begin{equation}\label{c14}
N=\frac{4\pi^{(D-1)/2}}{\Gamma\left(\frac{D-1}{2}\right)}\int_0^Rdr\theta\sigma^2r^{D-2}
\sqrt{\frac{B}{A}}\ .
\end{equation}
The total electric charge is $Q=eN$.

\begin{figure}
\centering
\includegraphics{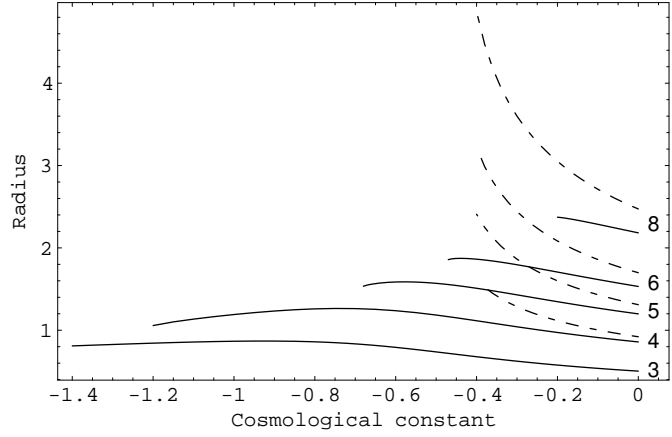}
\caption{The radius of a q-star as a function of the cosmological
constant. In figures \ref{figure8}-\ref{figure14} we take
$A_{\textrm{sur}}=0.81$, equivalently
$\omega=\theta_{\textrm{sur}}=0.45$.} \label{figure8}
\end{figure}

\begin{figure}
\centering
\includegraphics{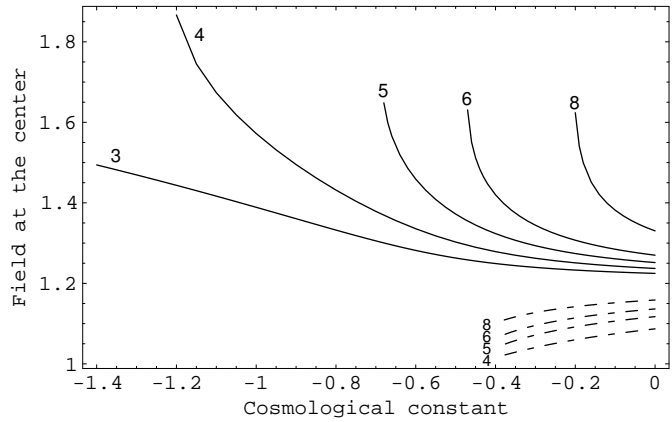}
\caption{The value of the scalar field at the center of the star
as a function of the cosmological constant.} \label{figure9}
\end{figure}

\begin{figure}
\centering
\includegraphics{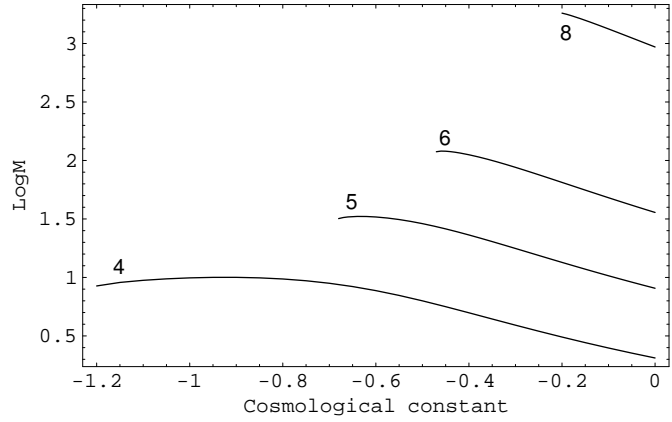}
\caption{The mass of a q-star with global symmetry as a function
of the cosmological constant.} \label{figure10}
\end{figure}

\begin{figure}
\centering
\includegraphics{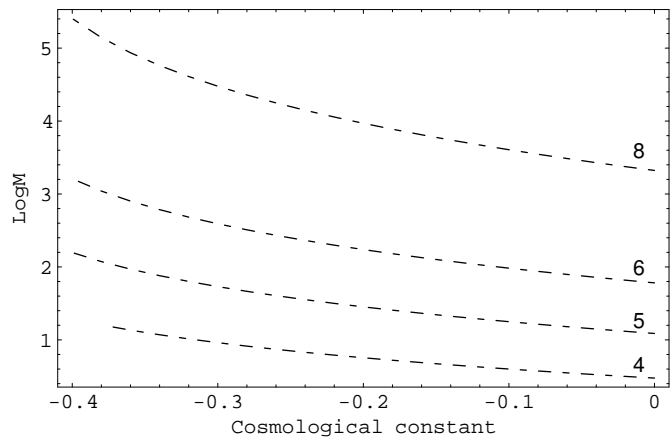}
\caption{The mass of a charged q-star with $e=1$ as a function of
the cosmological constant.} \label{figure11}
\end{figure}

\begin{figure}
\centering
\includegraphics{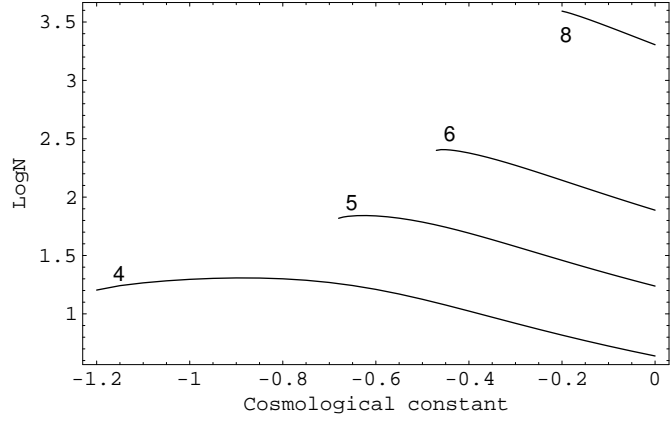}
\caption{The particle number of a q-star with global symmetry as
a function of the cosmological constant.} \label{figure12}
\end{figure}

\begin{figure}
\centering
\includegraphics{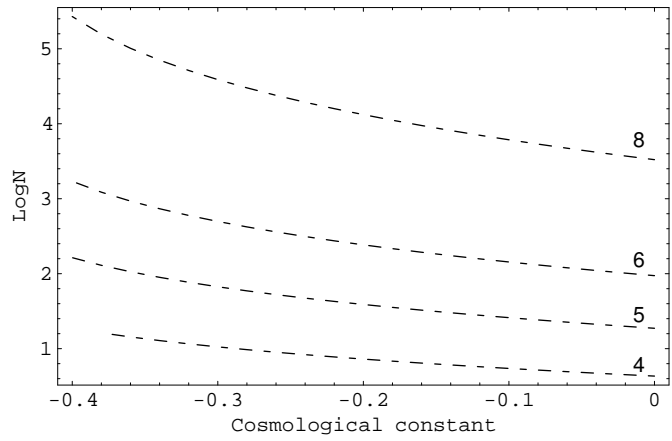}
\caption{The particle number of a charged q-star with $e=1$ as a
function of the cosmological constant.} \label{figure13}
\end{figure}

\begin{figure}
\centering
\includegraphics{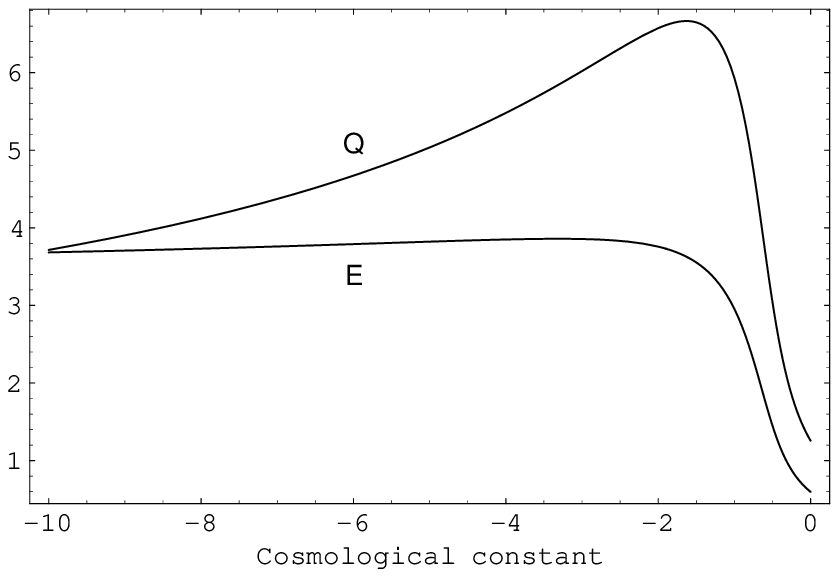}
\caption{The mass and the particle number of a q-star in $2+1$
dimensions as a function of the cosmological constant.}
\label{figure14}
\end{figure}

\section{Conclusions}

We find stable with respect to fission into free particles
q-star-type field configurations in 3, 4, 5, 6, 8 and 11
dimensions. We investigate the overall phase space. The
independent parameters are the frequency, $\omega$ or
$\theta_{\textrm{sur}}$, the cosmological constant and the field
strength, $e$. The star radius, the particle number, the total
electric charge, $eN$, and the mass of the star are derivative
quantities. The star shows a similar behavior in $D\geq4$
dimensions. Its behavior in $2+1$ seems to be quite different.
Charged q-stars in $2+1$ dimensions are not considered. The
numbers within the figures denote the spacetime dimensionality.
Dashed lines correspond to charged q-stars with $e=1$. Figures
\ref{figure1}-\ref{figure7} refer to asymptotically flat
spacetime and figures \ref{figure8}-\ref{figure14} refer to
asymptotically anti de Sitter spacetime. All the field
configurations depicted in our figures are stable with respect to
fission into free particles.

In figures \ref{figure1}-\ref{figure7} the frequency, $\omega$ or
$\theta_{\textrm{sur}}$ for the charged case, is the independent
parameter. We start form $\omega=\theta_{\textrm{sur}}=1/2$,
equivalently $A_{\textrm{sur}}=1$, which corresponds to the
absence of gravity (q-ball limit). We gradually decrease the
frequency down to a minimum value, determined by the soliton
stability demand. For the case of global symmetry, the smaller
(or critical) frequency corresponds to the last stable field
configuration with respect to gravitational collapse. (We also
include a short range of frequencies smaller than the critical
one.) In figure \ref{figure7} the last stable configuration with
respect to gravitational collapse corresponds to the top of the
$M=M(R)$ curve. For charged q-stars the lowest frequency
corresponds to $M/N=1$. Below this frequency the star mass is
larger than the mass of the free particles with the same charge,
and the star decays into free particles. No gravitational
collapse can happen when $e\neq0$. In general, charged q-stars
have larger radii and masses but a smaller $\phi(0)$, due to the
electrostatic repulsion between the different parts of the star.
They avoid for the same reason gravitational collapse, because for
$A_{\textrm{sur}}\rightarrow0$, equivalently
$\theta_{\textrm{sur}}\rightarrow0$, the electrostatic repulsion
makes the fission into free particles energetically favorable.

Figures \ref{figure8}-\ref{figure14} refer to spacetime with
negative cosmological constant. The independent parameter is the
cosmological constant $\Lambda$. We start from $\Lambda=0$ and
gradually decrease its value. We interrupt calculations when
$\phi(0)\rightarrow\infty$ for $e=0$ and when $M/N$ tends to
unity for a charged q-star.

The behavior of the star parameters, $R$, $M$, $N$ and $\phi(0)$,
for 4, 5 and 6 dimensions as a function of the cosmological
constant is similar. The situation is absolutely different in
$2+1$ dimensions, as we see from figure \ref{figure14}. For small
values of the cosmological constant both $N$ and $M$ increase
with the increase in absolute values of $\Lambda$. Below
$\Lambda\simeq-1.6$ the energy is approximately constant but the
particle number decreases slowly and for $\Lambda\simeq-10$
equals to the energy. Below that value of the cosmological
constant the star is unstable, decaying to free particles.

In four or more spacetime dimensions the total energy, particle
number and radius increases initially with the increase of the
cosmological constant in absolute values, but below a certain
value of the cosmological constant, these parameters decrease.
Qualitatively, negative cosmological constant reflects a
competitive effect to gravity attraction. So, the soliton mass
and consequently the particle number and radius increase, in
order the field configuration to be stable against this
``negative" gravity implied by the negative cosmological
constant. But, when $\Lambda$ exceeds a certain value, no
additional energy amount can deserve the soliton stability,
\textit{if this is too extended}. So, when $\Lambda$ exceeds this
certain value, the star shrinks, and, consequently, its energy
and charge decreases. Also, the value of the scalar field at the
center of the soliton increases rapidly for the same reason. This
happens because gravity in $D>3$ dimensions is a long-range force.
The two independent Einstein equations in static, spherically
symmetric systems contain the $(A-1)(D-3)(D-2)/(2r^2)$ term. On
the other hand, in $2+1$ dimensions no such competition between
``negative" gravity from the cosmological constant and
``positive" gravity from the energy-momentum density happens,
because gravity in $2+1$ dimensions is not a propagating
long-range force. Also, the above mentioned term is absent. These
are the reasons for the similarities in the behavior of the
soliton parameters in $D>3$ dimensions and the differences in
$2+1$ dimensions.

\vspace{1em}

\textbf{ACKNOWLEDGEMENTS}

\vspace{1em}

I wish to thank N.D. Tracas, E. Papantonopoulos and P.
Manousselis for helpful discussions.


\begin{thebibliography}{04}
\bibitem{boson-stars-old1}D. J. Kaup, Phys. Rev. \textbf{172}, 1331 (1968).
\bibitem{boson-stars-old2}R. Ruffini and S. Bonazzola, Phys. Rev. \textbf{187}, 1767
(1969).
\bibitem{boson-stars-new}E. W. Mielke and R. Scherzer, Phys. Rev. D \textbf{24}, 2111 (1981).
\bibitem{boson-stars-quartic}M. Colpi, S. L. Shapiro, and I. Wasserman,
Phys. Rev. Lett. \textbf{57}, 2485 (1986).
\bibitem{boson-stars-charged}Ph. Jetzer and J. J. van der Bij, Phys. Lett. B
\textbf{227}, 341 (1989).
\bibitem{nontop-solitons}T. D. Lee and Y. Pang, Phys. Rep.
\textbf{221}, 251 (1992).
\bibitem{large-soliton-stars1}R. Friedberg,
T. D. Lee, and Y. Pang, Phys. Rev. D \textbf{35}, 3640 (1987).
\bibitem{large-soliton-stars2}R. Friedberg, T. D. Lee, and Y. Pang, Phys. Rev. D
\textbf{35}, 3658 (1987).
\bibitem{large-soliton-stars3}R. Friedberg, T. D. Lee, and Y. Pang, Phys. Rev. D
\textbf{35}, 3678 (1987).
\bibitem{qballs-initial}S. Coleman, Nucl. Phys. \textbf{B262}, 263
(1985).
\bibitem{qballs-nonabelian}A. M. Safian,
S. Coleman, and M. Axenides, Nucl. Phys. \textbf{B297}, 498
(1988).
\bibitem{qballs-charged}K. Lee, J. A. Stein-Schabes, R. Watkins, and L. M.
Widrow, Phys. Rev. D \textbf{39}, 1665 (1989).
\bibitem{gibbons}G. W. Gibbons and S. W. Hawking, Phys. Rev. D
\textbf{15}, 2752 (1977).
\bibitem{qstars-global}B. W. Lynn, Nucl. Phys. \textbf{B321}, 465 (1989).
\bibitem{qstars-nonabelian}S. B. Selipsky, D. C. Kennedy and B. W. Lynn, Nucl. Phys.
\textbf{B321}, 430 (1989).
\bibitem{qstars-fermion}S. Bahcall, B. W. Lynn, and S. B. Selipsky, Nucl. Phys.
\textbf{B325}, 606 (1989).
\bibitem{qstars-charged}A. Prikas, Phys. Rev. D \textbf{66}, 025023 (2002).
\bibitem{qstars-AdS}A. Prikas, hep-th/0403019.
\bibitem{bosonstar-scalartensor-charged}
A. W. Whinnet and D. F. Torres, Phys. Rev. D \textbf{60}, 104050
(1999).
\bibitem{witten}E. Witten, Adv. Theor. Math. Phys. \textbf{2}, 253
(1998).
\bibitem{maldacena}J. M. Maldacena, Adv. Theor. Math. Phys.
\textbf{2}, 231 (1998).
\bibitem{string review}O. Aharony, S. S. Gubser, J. M. Maldacena, H.
Ooguri, and Y. Oz, Phys. Rept. \textbf{323}, 183 (2000).
\bibitem{boson-stars-extradim}D. Astefanesei and E. Radu, gr-qc/0309131.
\end{thebibliography}
\end{document}